\documentclass[aps,prd,twocolumn,groupedaddress,amsmath,amssymb]{revtex4-1}
\usepackage{graphicx}  
\usepackage{dcolumn}   
\usepackage{bm}        
\usepackage{verbatim}   
\usepackage[utf8]{inputenc}
\usepackage{color}

\begin{document}

\title{Quarkonium polarization in Pb+Pb collisions in the improved color evaporation model}
\author{Vincent Cheung}
\affiliation{
   Nuclear and Chemical Sciences Division,
   Lawrence Livermore National Laboratory,
   Livermore, California 94551, USA
   }
\author{Ramona Vogt}
\affiliation{
   Nuclear and Chemical Sciences Division,
   Lawrence Livermore National Laboratory,
   Livermore, California 94551, USA
   }
\affiliation{
   Department of Physics,
   University of California, Davis,
   Davis, CA 95616, USA
   }

\begin{abstract}
We apply our calculation of direct $J/\psi$ production in the improved color evaporation model at ${\mathcal O}(\alpha_s^3)$ in collinear factorization to Pb+Pb collisions.  The cold nuclear matter effects of nuclear modifications of the parton densities and intrinsic transverse momentum broadening are included.  We find that the polarization is independent of these effects.  Our calculations are compared to the Pb+Pb polarization parameters extracted by the ALICE Collaboration at $\sqrt{s_{NN}} = 5.02$~TeV.
\end{abstract}


\maketitle


\section{Introduction}

Quarkonium production provides important insights into hadronization in QCD.  Quarkonium has been studied in a variety of systems: $e^+ + e^-$, $e + p$, $\gamma + \gamma$, $p + \overline p$, $p+p$, $p+A$ and $A+A$ collisions.  Although suppression of quarkonium in $A+A$ collisions was proposed to be the definite signature of quark-gluon plasma formation \cite{MatsuiSatz}, the original proposition did not take into account the fact that quarkonium production was already modified by cold nuclear matter effects in $p+A$ collisions.  Indeed, quarkonium production has yet to be fully understood in the more elementary collisions, including $p+p$ and $p + \overline p$ collisions.

Several approaches have been proposed for quarkonium production in $p+p$ collisions: the color singlet model \cite{BaierRuckl}, the color evaporation model (CEM) \cite{Barger:1979js,Barger:1980mg,Gavai:1994in,Ma:2016exq}, and the effective field theory, nonrelativistic QCD (NRQCD) \cite{Caswell:1985ui,Bodwin:1994jh} among them.  Given that both NRQCD and the CEM, including the recent improved CEM (ICEM) \cite{Ma:2016exq}, can describe the yields, including the transverse momentum, $p_T$, distributions, other means of distinguishing between models are required.

The polarization of quarkonium production was proposed as a means of distinguishing between model approaches.  While it has been measured for quarkonium production in hadron colliders at high $p_T$ \cite{CMSpol}, as well as in fixed-target experiments at low $p_T$ \cite{NuSea:2003km,Biino:1987qu}, it has proved difficult to describe both the yields and polarizations simultaneously within a given approach \cite{Nora}.

A comparison of the polarization in $p+p$ and Pb+Pb collisions is relevant because $J/\psi$ production is strongly suppressed in Pb+Pb collisions relative to $p+p$.  Feed down production of $J/\psi$ from the higher mass quarkonium states $\chi_{cJ}$ and $\psi$(2S) is assumed to be suppressed first.  At low $p_T$, particularly at midrapidity, there is a significant contribution to $J/\psi$ production from the regeneration of $J/\psi$ from uncorrelated $c \overline c$ pairs in the medium.  After production, as the ``proto"-$J/\psi$ moves through the medium, color singlets and color octets could have different suppression rates.  Any of these effects due to the presence of the hot medium could potentially change the polarization in Pb+Pb collisions relative to the $p+p$ baseline.

The Pb+Pb measurements could provide a strong constraint on the $J/\psi$
production mechanism.  If $J/\psi$ hadronization is a fast process, effectively evaporating color, as in the ICEM, the polarization should not change significantly in Pb+Pb collisions relative to $p+p$.  If, on the other hand, $J/\psi$ production involves distinct contributions in color singlet and color octet states, as in NRQCD, and these states have different suppression patterns in the medium \cite{NoraEtAl}, the polarization should depend on the QCD medium in which the $J/\psi$ is formed and hadronizes.  The only circumstance under which this might not be the case would be if hadronization takes place far outside the nucleus in both color states.  In NRQCD, the relative values of these color octet states, fixed by fits of the long-distance matrix elements, determine the polarization.  At high $p_T$, unless otherwise fit to polarization data, the polarization parameter, $\lambda_\vartheta$, suggests that $J/\psi$ production is transversely polarized (positive $\lambda_\vartheta$).  As shown in Ref.~\cite{Faccioli:2014ca} where $\lambda_\vartheta$ was calculated separately for each color and spin state, the $^1$S$_0^{[8]}$ state is unpolarized, the color singlet contribution is longitudinally polarized for $p_T > 10$~GeV, and the $^3$S$_1^{[8]}$ state is transversely polarized for $p_T > 10$~GeV.  The $^3$P$_J^{[8]}$ was excluded from the analysis.  In the ICEM, the polarization is rather weak, with a slight longitudinal polarization at low $p_T$, and then either unpolarized or slightly transversely polarized at high $p_T$, depending on the polarization frame. \cite{CheungVogt_coll}.

Experimentally, in $p+p$ collisions it has been observed that $\lambda_\vartheta \approx 0$.  The first measurement of the quarkonium polarization in Pb+Pb collisions at $\sqrt{s_{NN}} = 5.02$~TeV was presented by the ALICE Collaboration \cite{ALICE:polPbPb} as a function of transverse momentum, $p_T$, in the range $2 < p_T < 15$~GeV in the forward rapidity region $2 < y < 4.5$.  The polarization parameters extracted in Pb+Pb collisions were found to be compatible with those measured in $p+p$ collisions within the uncertainties.  The polarization parameters were previously measured by ALICE in $p+p$ collisions at $\sqrt{s} = 7$ and 8~TeV.

Here, we study $J/\psi$ polarization in Pb+Pb collisions in the ICEM assuming only cold nuclear matter effects on $J/\psi$ production.  The production and polarization of $J/\psi$ in the ICEM in $p+p$ collisions are described in Sec.~\ref{Sec:ICEM_pol}.  Modifications due to the presence of a cold nuclear medium in Pb+Pb collisions are taken into account in Sec.~\ref{Sec:ICEM_AApol}.  The resulting calculations are compared to the $p+p$ and Pb+Pb data in Sec.~\ref{Sec:Results}, followed by a discussion and summary in Sec.~\ref{Sec:Summary}.


\section{Polarization in the ICEM}
\label{Sec:ICEM_pol}

The ICEM assumes the $J/\psi$ production cross section is a constant fraction of the open $c\bar{c}$ cross section with invariant mass above the $J/\psi$ mass but below the charm hadron pair mass threshold, the mass of two $D^0$ mesons. In the ICEM, distinction is also made between the momentum of the $c\bar{c}$ pair and that of the $J/\psi$, compared to the traditional CEM where no difference between the two is assumed. The unpolarized direct $J/\psi$ production cross section in $p+p$ collisions in the ICEM is
\begin{eqnarray}
\label{ch6-icem-cross-section}
\sigma &=& F_{J/\psi} \sum_{i,j}  \int^{2m_D}_{M_{J/\psi}}dM dx_i dx_j f_i(x_i,\mu_F)f_j(x_j,\mu_F) \nonumber \\
&\times& \hat{\sigma}_{ij\rightarrow c\bar{c}+k}(p_{c\bar{c}},\mu_R) |_{p_{c\bar{c}} = \frac{M}{M_{J/\psi}} p_{\psi}} \;,
\end{eqnarray}
where $i$ and $j$ are $q, \bar{q}$ and $g$ with $ij = q\bar{q}$, $qg$, $\bar{q}g$, or $gg$.  Here $F_{J/\psi}$ is a universal factor at fixed order for direct $J/\psi$ production in the ICEM, independent of the projectile, target, and center-of-mass energy, $x$ is the momentum fraction of the parton, and $f(x,\mu_F)$ is the parton distribution function (PDF) in the proton as a function of $x$ and the factorization scale $\mu_F$. Finally, $\hat{\sigma}_{ij\rightarrow c\bar{c}+k}$ are the partonic cross sections for the initial state $ij$ to produce a $c\bar{c}$ pair with a light final-state parton $k$. The amplitudes for the partonic cross sections were described in detail in Ref.~\cite{CheungVogt_coll}.  The invariant mass of the $c\bar{c}$ pair, $M$, is integrated from the physical mass of $J/\psi$ ($M_{J/\psi} =3.10$~GeV) to twice the $D^0$ mass ($2m_{D^0} = 3.72$~GeV).

The angular momentum of the proto-$J/\psi$ is assumed to be unchanged by hadronization of the $c \overline c$ pair to the final-state $J/\psi$, consistent with the philosophy of the color-evaporation model \cite{Barger:1979js,Barger:1980mg}.  To obtain the $J/\psi$ cross section, the partonic cross sections are convoluted with the CT14 PDFs \cite{Dulat:2015mca} for $p_\psi \cdot k =0$. The partonic cross section calculations are restricted to be within the perturbative domain by introducing a regularization parameter requiring that all propagators are at a minimum distance of $Q_{\rm reg}^2=M^2$ from their poles, as employed in Ref.~\cite{Baranov:2002cf}. The factorization and renormalization scales are taken to be $\mu_F/m_T = 2.1^{+2.55}_{-0.85}$ and $\mu_F/m_T = 1.6^{+0.11}_{-0.12}$ respectively, where $m_T$ is the transverse mass of the charm quark produced ($m_T = \sqrt{m_c^2+p_T^2}$ and $p_T^2 = 0.5\sqrt{p_{Tc}^2+p_{T\bar{c}}^2}$). The charm quark mass is varied around the central value as $1.27\pm 0.09$~GeV. The mass and scale central values and their uncertainties were determined in Ref.~\cite{Nelson:2012bc} to obtain the uncertainty on the total charm cross section.

Because the $\mathcal{O}(\alpha_s^3)$ contribution diverges when the light parton is soft, the initial-state partons are given a small transverse momentum, $k_T$, kick to describe the $p_T$ distribution at low $J/\psi$ $p_T$. This kick is assumed to be $\langle k_T^2\rangle = 1+ (1/12) \ln(\sqrt{s}/20 {\rm ~GeV})$ or $1.49{\rm ~GeV}^2$ for $\sqrt{s}=7$~TeV. The collinear PDFs are multiplied by a Gaussian function $g(k_{T})$,
\begin{eqnarray}
g(k_T) &=& \frac{1}{\pi \langle k_T^2 \rangle} \exp (k_T^2/\langle k_T^2\rangle) \;, \label{Eq:gofkT}
\end{eqnarray}
assuming the $x$ and $k_T$ dependences factorize. The same Gaussian smearing is applied in Refs.~\cite{Nelson:2012bc,Ma:2016exq,Mangano:1991jk,MLM1}. Note that in the traditional CEM, the lower invariant mass threshold for all charmonium states is set to the production threshold, which makes the kinematic distributions of the charmonium states, $\mathcal{Q}$, identical except for the choice of $F_{\mathcal{Q}}$. The distinction between the $J/\psi$ and $c\bar{c}$ momenta in Eq.~(\ref{ch6-icem-cross-section}) also helps describe the $p_T$ distributions at high $p_T$.

The amplitudes calculated in Ref.~\cite{CheungVogt_coll} can be factorized so that the polarization vector, $\epsilon_\psi(J_z)$, is extracted from the unsquared amplitudes for all sub-processes, giving
\begin{eqnarray}
\mathcal{M}_n &=& \epsilon_\psi^\mu(J_z) \mathcal{M}_{n,\mu}
\end{eqnarray}
for each subprocess denoted by the initial states, $n=gg,gq,g\bar{q},q\bar{q}$. The polarization vectors for $J_z=0$, $\pm1$ in the rest frame of the proto-$J/\psi$ are
\begin{eqnarray}
\epsilon_\psi(0)^\mu &=& (1,0,0,0) \;, \\
\epsilon_\psi(\pm1)^\mu &=& \mp\frac{1}{\sqrt{2}}(0,1,\pm i,0) \;,
\end{eqnarray}
with the convention that the $z$ direction is the fourth component of the four-vector.  Although the unpolarized cross section is independent of the choice of $z$ axis, the polarized cross sections depend on the orientation of the $z$ axis. In these calculations, the $y$ axis is chosen to be the vector normal to the plane formed by the two beams with momenta $\vec{P}_1$ and $\vec{P}_2$,
\begin{eqnarray}
\hat{y} &=& \frac{-\vec{P}_1\times\vec{P}_2}{|\vec{P}_1\times\vec{P}_2|} \;.
\end{eqnarray}
Two orientations of the $z$ axis are considered: the helicity frame, HX, and the Collins-Soper frame, CS.
In the helicity frame, $z_{\rm HX}$ is the flight direction of the $c\bar{c}$ pair in the center of mass of the colliding beams. In the Collins-Soper frame \cite{Collins:1977iv}, $z_{CS}$ is the angle bisector between one beam and the opposite direction of the opposing beam. In both cases, the $x$ axis is determined by the right-handed convention.

Although the unpolarized cross section is a single value, averaged over all polarization directions, the polarization parameters are described in terms of matrix elements of the polarized cross section.  A given polarized cross section matrix element, $\sigma_{i_z,j_z}$, is calculated in the rest frame of the $c\bar{c}$ pair as follows.  First, we take the product of the unsquared amplitude with polarization vector $J_z=i_z$ and the unsquared amplitude with polarization vector $J_z=j_z$ in each subprocess, denoted by the initial states ($n = gg$ ,$gq$, $g\overline{q}$), $q\overline{q}$.  These are then added and the components of the polarized cross section matrix are calculated employing Eq.~(\ref{ch6-icem-cross-section}),
\begin{eqnarray}
\sigma_{i_z,j_z} &=& \int \sum_{n} (\epsilon_\psi^\mu(i_z) \mathcal{M}_{n,\mu})(\epsilon_\psi^{\nu}(j_z) \mathcal{M}_{n,\nu})^* \;,
\label{Eq:pol_sigma}
\end{eqnarray}
where $i_z,j_z=\{-1,0,+1\}$. The integral in Eq.~(\ref{Eq:pol_sigma}) is over all variables explicitly shown in Eq.~(\ref{ch6-icem-cross-section}) as well as the Lorentz-invariant phase space for $2\rightarrow3$ scatterings.

The unpolarized cross section is the trace of the polarized cross section matrix
\begin{eqnarray}
\sigma_{\rm unpol} &=& \sum_{i_z} \sigma_{i_z,i_z} = \sigma_{-1,-1} + \sigma_{0,0} + \sigma_{+1,+1} \;.
\end{eqnarray}

The polarization parameters are calculated using the polarized matrix elements given Eq.~(\ref{Eq:pol_sigma}). The polar anisotropy, $\lambda_{\vartheta}$, the azimuthal anisotropy, $\lambda_\varphi$, and the polar-azimuthal correlation, $\lambda_{\vartheta\varphi}$, are defined as \cite{Faccioli:2010kd}
\begin{eqnarray}
\lambda_{\vartheta} &=& \frac{\sigma_{+1,+1}-\sigma_{0,0}}{\sigma_{+1,+1}+\sigma_{0,0}} \; \label{lambda_theta_eqn} ,\\
\lambda_{\varphi} &=& \frac{\operatorname{Re}[\sigma_{+1,-1}]}{\sigma_{+1,+1}+\sigma_{0,0}} \;, \\
\lambda_{\vartheta\varphi} &=& \frac{\operatorname{Re}[\sigma_{+1,0}-\sigma_{-1,0}]}{\sqrt{2}(\sigma_{+1,+1}+\sigma_{0,0})} \label{lambda_theta_phi_eqn}\;.
\end{eqnarray}
These parameters depend on the frame (HX or CS) in which they are calculated and measured. Because the angular distribution is rotationally invariant, it is possible to construct invariant polarization parameters from Eqs.~(\ref{lambda_theta_eqn})–(\ref{lambda_theta_phi_eqn}). One frame-invariant polarization parameter is $\tilde{\lambda}$ \cite{Faccioli:2010kd}
\begin{eqnarray}
\tilde{\lambda} &=& \frac{\lambda_\vartheta+3\lambda_\varphi}{1 - \lambda_\varphi} \;.
\end{eqnarray}
The choice of $\tilde{\lambda}$ is the same as the polar anisotropy parameter $\lambda_\vartheta$ in a frame where the distribution is azimuthally isotropic $\lambda_{\varphi}=0$. It is possible to remove the frame-induced kinematic dependencies when comparing theoretical predictions to data by considering $\tilde{\lambda}$.

\section{Nuclear Matter Effects in Pb+Pb Collisions}
\label{Sec:ICEM_AApol}

There are a number of cold nuclear matter effects important for $J/\psi$ production.  These can include transverse momentum, $k_T$, broadening; modification of the parton distributions in the nucleus (nPDFs); nucleon absorption; and dissociation by comovers.  Nucleon absorption is expected to be negligible at LHC energies.  Suppression by comovers is neglected here but nPDF effects and $k_T$ broadening are included.

The EPPS16 modifications of the parton densities in the nucleus \cite{EPPS16} are incorporated into the calculations.  This set includes LHC data taken at $\sqrt{s_{NN}} = 5.02$~TeV, in particular dijet production which provides information on the gluon modification in the nucleus and gauge boson production which provides constraints on the nuclear quark and antiquark distributions.

The intrinsic $\langle k_T^2 \rangle$ used in Eq.~(\ref{Eq:gofkT}) is expected to be broadened due to the presence of the nucleus---see Refs.~\cite{RV_azi1,RV_azi2,RV_SeaQuest}---so that $\langle k_T^2 \rangle_A = \langle k_T^2 \rangle_p + \delta k_T^2 = 1.9$~GeV$^2$ since the intrinsic $k_T$ kick in a lead nucleus is taken to be $\delta k_T^2 = 0.41$~GeV$^2$.  In Pb+Pb collisions, this kick is added to partons from both nuclei.

In nucleus-nucleus collisions, however, additional effects can be important because of the hot matter created in these collisions.  In this case, the hot matter can be expected to almost completely suppress the production of some quarkonium states.  For example, the temperature of the hot medium is sufficient to suppress $\chi_c$ and $\psi$(2S) production since these states are more loosely bound. Thus feed down contributions to inclusive $J/\psi$ production are effectively eliminated or strongly suppressed in central collisions.  The absence of feed down contributions to inclusive $J/\psi$ production can affect the polarization \cite{CheungVogt2}.

In addition, quarkonium states, once suppressed, can be regenerated in the hot medium because the density of $c$ and $\overline c$ quarks is high in nucleus-nucleus collisions \cite{Thews,Andronic}. Recent ALICE data \cite{ALICE_PbPb_centrality} show that, in the most central collisions, the nuclear modification factor $R_{AA}$ is $\approx 1$ at $p_T \approx 0$ at midrapidity, $|y| < 0.9$, while it is reduced to $\approx 0.8$ at $2.5 < y < 4$ for the same $p_T$.  The suppression factor decreases with $p_T$ until $p_T \approx 5$~GeV, at which point $R_{AA}(p_T) \approx 0.25$ independent of $p_T$ and rapidity.  In more peripheral collisions, at forward rapidity, $R_{AA}$ is reduced for $p_T < 5$~GeV and increased for higher $p_T$ until, for the 40-90\% centrality bin, $R_{AA} \approx 0.6$ over all $p_T$. These measurements show that, as predicted, regeneration of the $J/\psi$ is concentrated at low $p_T$ and more important at midrapidity than at forward rapidity.  Regeneration is also shown to effectively disappear in more peripheral collisions.

Our approach has been shown to describe the $J/\psi$ $p_T$ distributions in Ref.~\cite{CheungVogt_coll}.  The $p+p$ $J/\psi$ polarization at $\sqrt{s} = 7$~TeV was also shown to be compatible with the $p_T$ dependence of the polarization parameters measured by the ALICE Collaboration \cite{ALICE:2011md}. (The polarization  at 8~TeV \cite{ALICE:2018crw} is compatible with that measured at 7~TeV.)  We now turn to the polarization in Pb+Pb collisions to see how the polarization changes when the cold nuclear matter effects of nuclear modifications according to EPPS16 and $k_T$ broadening are introduced.  We have not introduced any hot matter effects in our calculations.

\section{Results}
\label{Sec:Results}

In this section, we compare our calculations in the ICEM to the Pb+Pb $J/\psi$ polarization data taken by the ALICE Collaboration \cite{ALICE:polPbPb}.  In both
$p+p$ and Pb+Pb collisions, the theoretical uncertainties are obtained by varying the charm quark mass, the renormalization scale, and the factorization scale as discussed in Sec.~\ref{Sec:ICEM_pol}.  The uncertainty band is constructed by adding the mass and scale uncertainties in quadrature.  In the Pb+Pb calculations, the nPDF uncertainties are also included.

\begin{figure*}
\centering
\begin{minipage}[ht]{0.68\columnwidth}
\centering
\includegraphics[width=\columnwidth]{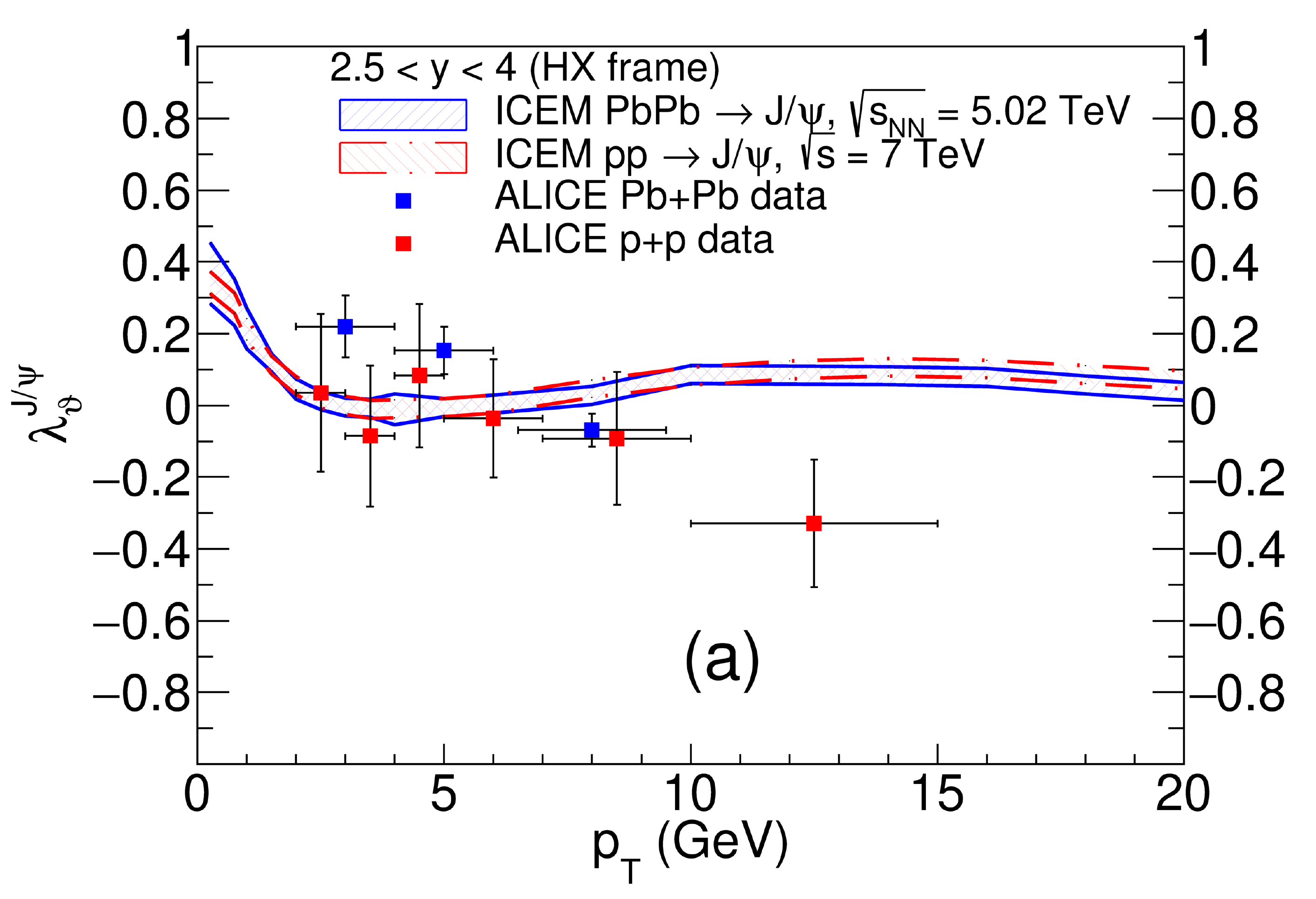}
\end{minipage}%
\begin{minipage}[ht]{0.68\columnwidth}
\centering
\includegraphics[width=\columnwidth]{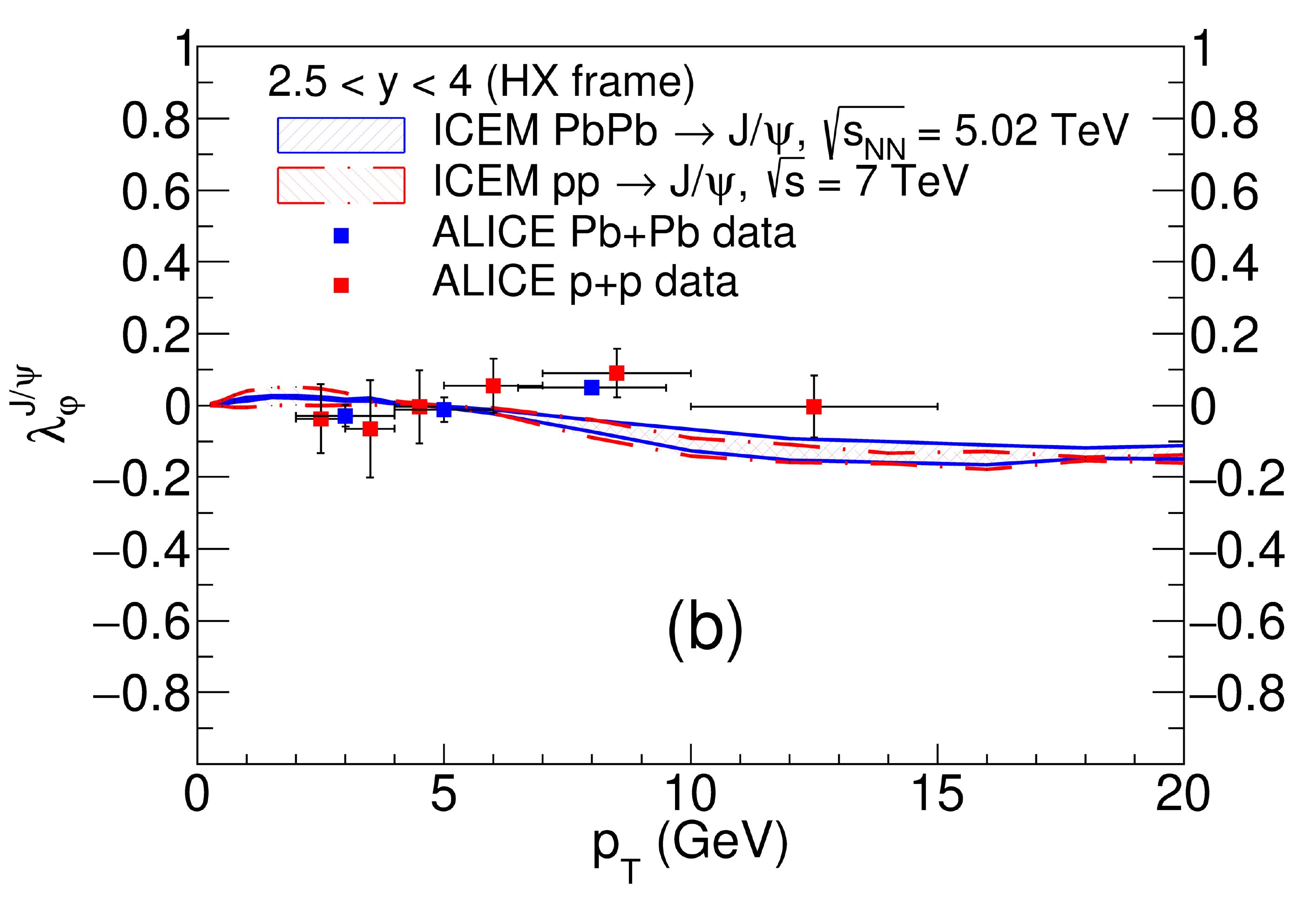}
\end{minipage}
\begin{minipage}[ht]{0.68\columnwidth}
\centering
\includegraphics[width=\columnwidth]{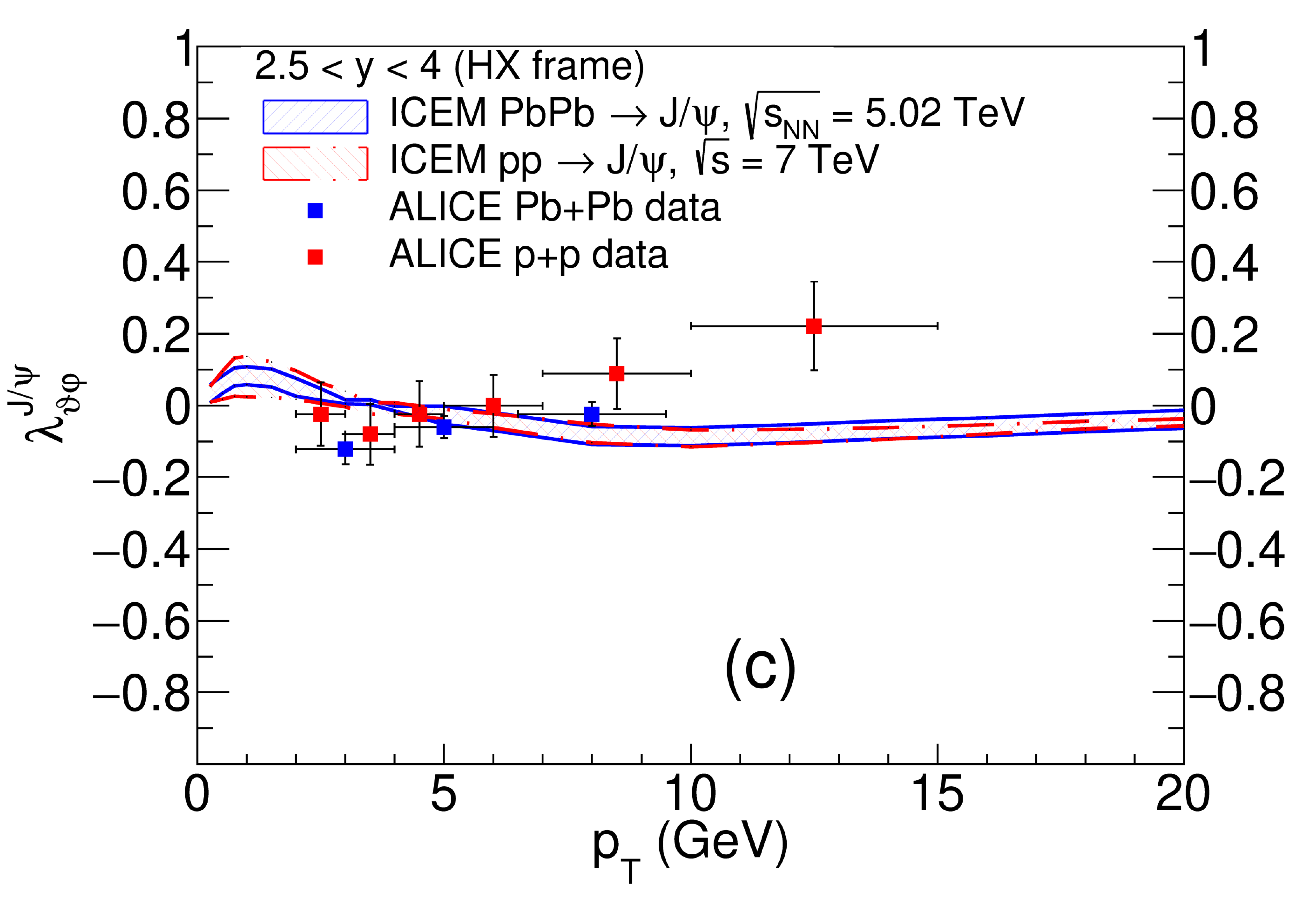}
\end{minipage}
\caption{(a) The polar anisotropy parameter ($\lambda_\vartheta$), (b) the azimuthal anisotropy parameter ($\lambda_\varphi$), and (c) the polar-azimuthal correlation parameter ($\lambda_{\vartheta\varphi}$) in the helicity frame in the ICEM. The combined mass, renormalization scale, and factorization scale uncertainties are shown in the band and compared to the ALICE Pb+Pb data at $\sqrt{s_{NN}} = 5.02$~TeV \cite{ALICE:polPbPb} (blue) and the ALICE $p+p$ data at $\sqrt{s} = 7$~TeV \cite{ALICE:2018crw} (red).} \label{frame-dependent-lambdas-hx}
\end{figure*}

\begin{figure*}
\centering
\begin{minipage}[ht]{0.68\columnwidth}
\centering
\includegraphics[width=\columnwidth]{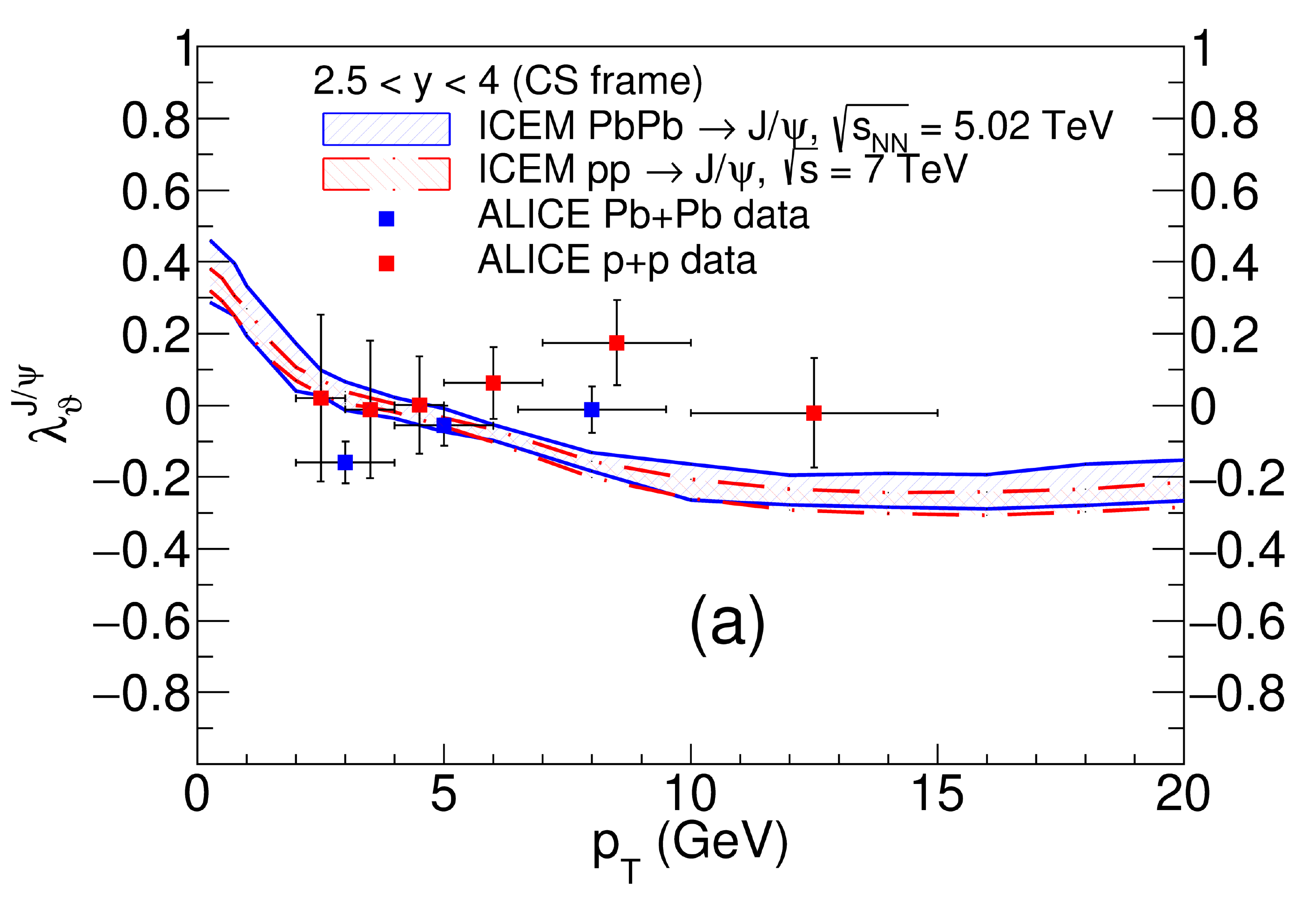}
\end{minipage}%
\begin{minipage}[ht]{0.68\columnwidth}
\centering
\includegraphics[width=\columnwidth]{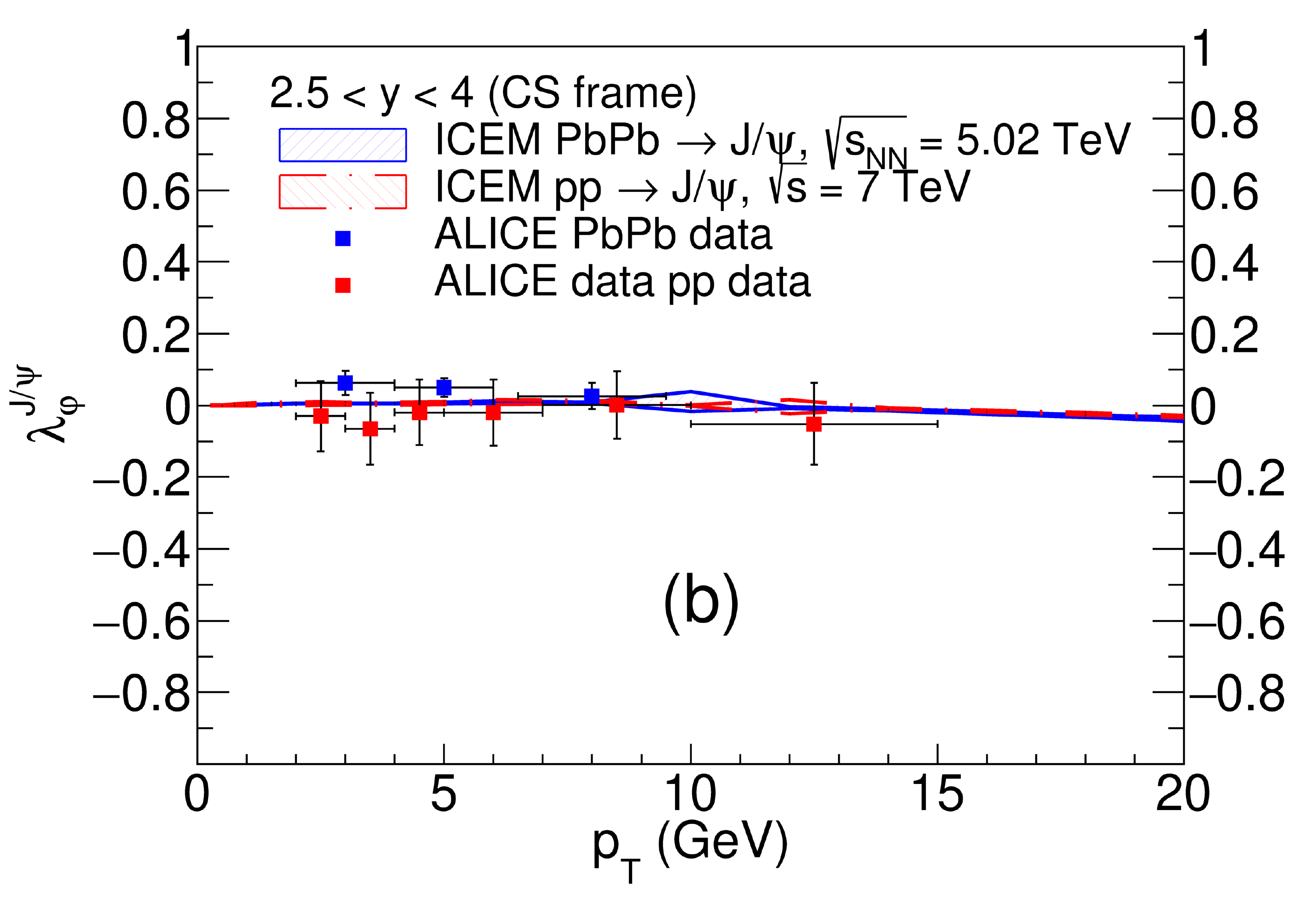}
\end{minipage}
\begin{minipage}[ht]{0.68\columnwidth}
\centering
\includegraphics[width=\columnwidth]{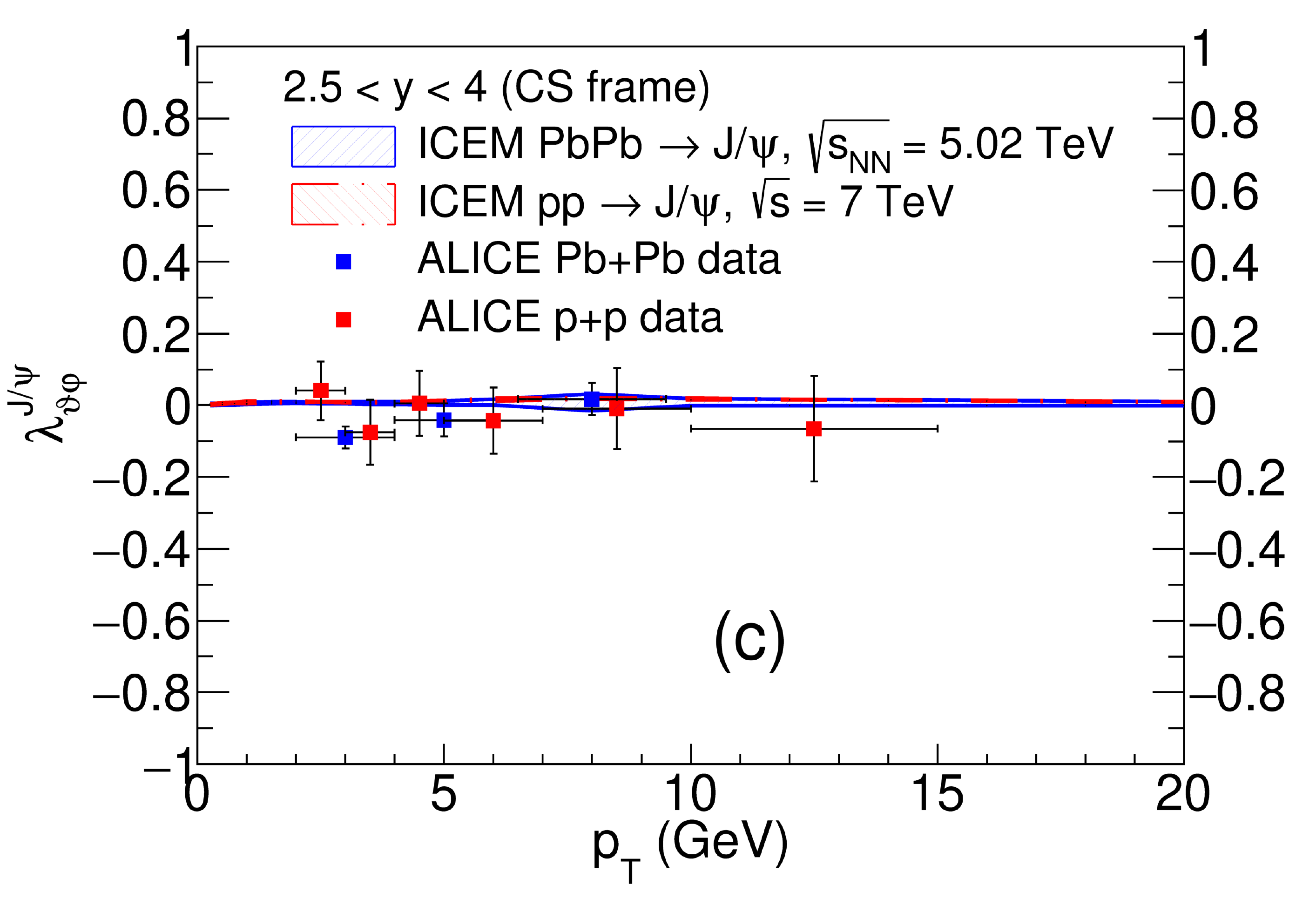}
\end{minipage}
\caption{(a) The polar anisotropy parameter ($\lambda_\vartheta$), (b) the azimuthal anisotropy parameter ($\lambda_\varphi$), and (c) the polar-azimuthal correlation parameter ($\lambda_{\vartheta\varphi}$) in the Collins-Soper frame in the ICEM. The combined mass, renormalization scale, and factorization scale uncertainties are shown in the band and compared to the ALICE Pb+Pb data at $\sqrt{s_{NN}} = 5.02$~TeV \cite{ALICE:polPbPb} (blue) and the ALICE $p+p$ data at $\sqrt{s} = 7$~TeV \cite{ALICE:2018crw} (red).} \label{frame-dependent-lambdas-cs}
\end{figure*}

\begin{figure}[h!]
\centering
\includegraphics[width=\columnwidth]{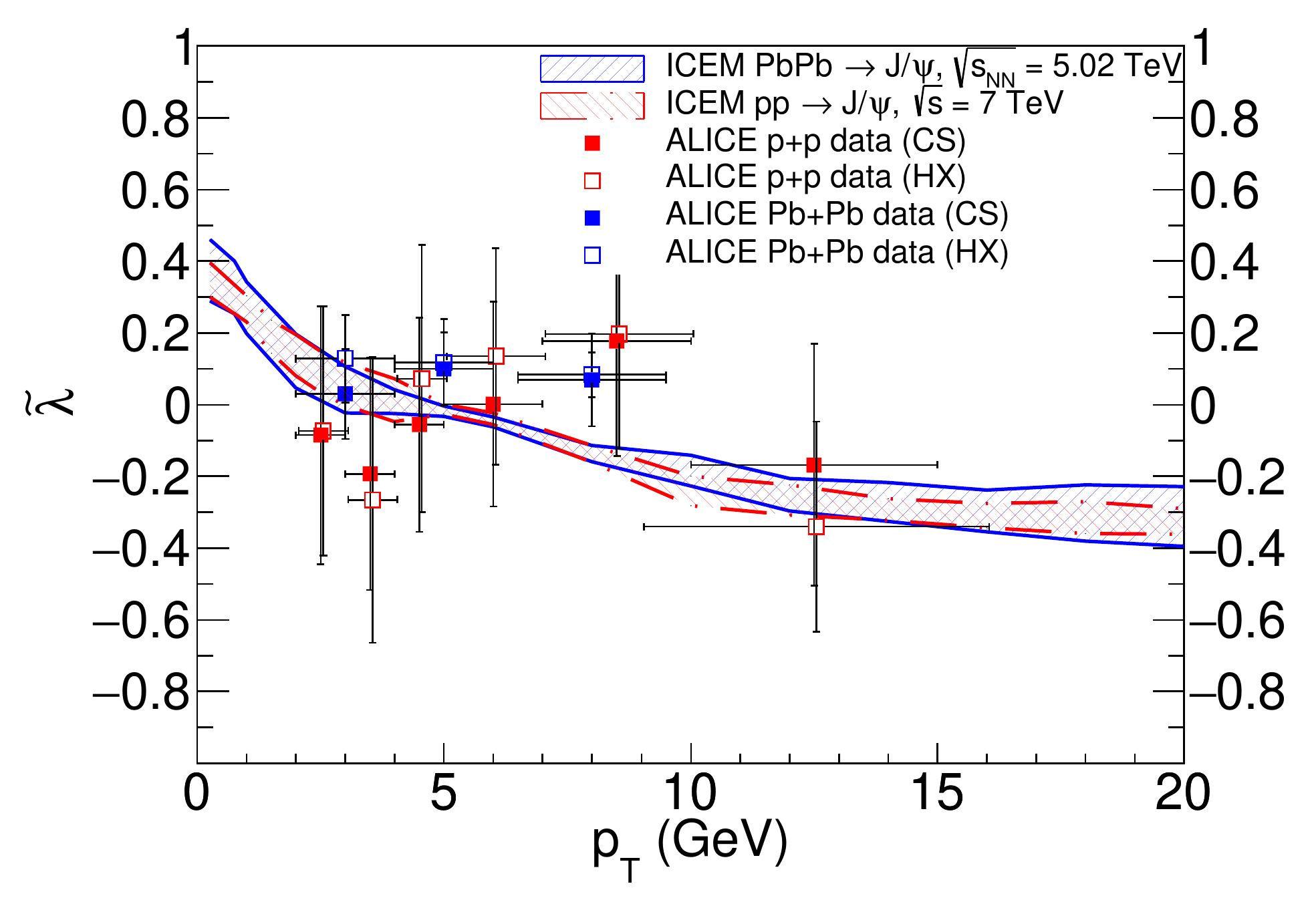}
\caption{The $p_T$ dependence of the frame-invariant polarization parameter, $\tilde{\lambda}$, in the ICEM compared to the ALICE Pb+Pb data \cite{ALICE:polPbPb} (blue) and the ALICE $p+p$ data \cite{ALICE:2018crw} (red). Data measured in the Collins-Soper frame are presented as solid points, and those in the helicity frame are presented as open points. The data in the helicity frame are displaced by 0.05~GeV for visualization purposes. } \label{lambda-invariant}
\end{figure}

Our calculations of the three polarization parameters defined in Eqs.~(\ref{lambda_theta_eqn})-(\ref{lambda_theta_phi_eqn}) are shown in Figs.~\ref{frame-dependent-lambdas-hx} and \ref{frame-dependent-lambdas-cs} for the helicity and Collins-Soper frames respectively.  The frame-invariant polarization parameter is shown in Fig.~\ref{lambda-invariant}.  The $p+p$ calculations are performed at $\sqrt{s} = 7$~TeV while the Pb+Pb calculations are made at $\sqrt{s_{NN}} = 5.02$~TeV.  The $p+p$ calculations include $k_T$ broadening as described in Eq.~(\ref{Eq:gofkT}) while the Pb+Pb calculations include an enhanced $k_T$ broadening in nuclear collisions using the EPPS16 central nPDF set.

No significant differences between the $p+p$ and Pb+Pb calculations are seen. In addition to the central EPPS16 set, we have also checked if there are any changes in the polarization parameters due to choosing another shadowing set.  The two error sets that exhibit the largest excursion from the central gluon set were used.  No difference in the results was seen.  We also increased the $k_T$ broadening employed and likewise saw no difference in the calculated polarization parameters.  The only slight difference in the calculated results is due to the 40\% difference in collision energy per nucleon.  A similar lack of system and energy dependence of the polarization is also expected in $p$+Pb collisions using the CGC+NRQCD approach \cite{Stebel:2021bbn}.

Despite the fact that differences in the yields have been observed as a function of $p_T$ due to cold nuclear matter effects such as shadowing and $k_T$ broadening, as exhibited in the nuclear modification factor, there should not be significant modifications of the calculated polarization in Pb+Pb
collisions relative to $p+p$ collisions, as discussed in
\cite{CheungVogt2,CheungVogt3}.  This is due to the way that the
polarization parameters are calculated: because they depend on the ratio of sums and differences of elements of the polarized cross section matrix, the numerator and denominator of the polarization parameters are affected similarly. As discussed in Ref.~\cite{CheungVogt3}, the polarization is only sensitive to is the heavy quark mass because the partonic cross sections depend directly on the mass.  Changes in the factorization and renormalization scales do not affect the amplitudes, only the evolution of the parton densities and the scale in $\alpha_s$, resulting in the polarization being effectively independent of the scales.  Thus, in $p+p$ calculations of the polarization, the uncertainty band is relatively narrow because all hot matter effects such as quarkonium regeneration in nuclear collisions are also unlikely to affect the polarization unless the various quarkonium spin states are regenerated in different proportions than in $p+p$ interactions.  So far such effects on regeneration have not been studied in detail.

Finally, we note here that the calculations are both for direct $J/\psi$ polarization, no feed down has so far been included.  In the quark-gluon plasma, sequential suppression of higher mass quarkonium states would lead to the expectation that there would be little to no feed down contribution to the $J/\psi$ polarization in Pb+Pb collisions due to suppression of the $\psi$(2S) and $\chi_c$ states.  In Ref.~\cite{CheungVogt3}, we investigated differences between direct and prompt $J/\psi$ polarization in the $k_T$ factorization approach and saw somewhat steeper slope in $\lambda_\vartheta$ for direct $J/\psi$ relative to prompt $J/\psi$ for $p_T > 8$~GeV.  The data in this region are subject to large uncertainties and are, so far, unable to discriminate between changes in polarization due to the potential loss of feed down from higher state suppression.


\section{Discussion and Conclusions}
\label{Sec:Summary}

The polarization data in Pb+Pb collisions shown in Figs.~\ref{frame-dependent-lambdas-hx} and \ref{frame-dependent-lambdas-cs} are minimum bias, 0-90\% centrality, including data over all collision centralities.  These minimum-bias data show that the polarization in the two systems is consistent within the measured statistical and systematic uncertainties.  The presence of the hot medium does not seem to affect the polarization, despite the fact that the medium effects strongly modify the yields through a combination of suppression and regeneration.

The small differences between the $p+p$ and Pb+Pb polarization imply that feed down from the excited charmonium states, or any lack thereof due to greater suppression of the excited states in the hot medium, does not strongly affect the prompt $J/\psi$ polarization.  Our previous results for direct and prompt $J/\psi$ in the $k_T$-factorization approach \cite{CheungVogt3} show that feed down can affect the polarization in $p+p$ collisions.  We will confirm if feed down affects the polarization in $p+p$ calculations the same way in the collinear factorization approach in a future publication.

The yields, studied as a function of $\sqrt{s_{NN}}$, centrality, $p_T$ and rapidity in recent ALICE measurements \cite{ALICE_PbPb_centrality}, show that regeneration effects on the $J/\psi$ $R_{AA}$ decrease as one goes from central to more peripheral collisions.  Regeneration is an important source of low $p_T$ $J/\psi$ in central collisions but becomes negligible in more peripheral events.  The source of the measured $J/\psi$ thus changes with collision centrality.  The difference in the results between $\sqrt{s_{NN}} = 2.76$ and 5.02~TeV was shown to be small.

If the hadronization of the regenerated $J/\psi$ is different than that of the initially-produced $J/\psi$, with different admixtures of color and spin states, then this difference could affect the measured polarization.   An analysis of the polarization at low $p_T$ could then demonstrate how much the polarization depends on the time the $J/\psi$ is produced (in the initial nucleon-nucleon collisions or from deconfined $c$ and $\overline c$ quarks in the hot medium created by the collision) and the density of the medium. If the regenerated $J/\psi$ are a mixture of color and spin states, then the polarization at low $p_T$ would be similar to that in $p+p$ collisions—nearly unpolarized—while, at higher $p_T$, it could become transversely polarized because of color octet suppression. There is, unfortunately, little discussion of the color and spin of the regenerated quarkonium states. However, the $R_{AA}$ can be described by transport models that include suppression and regeneration effects in the medium \cite{DuRapp}. These models can also describe the centrality dependence of $R_{AA}$ as measured by ALICE \cite{ALICE_PbPb_centrality}.

The centrality dependence of the polarization can reveal if there are strong differences in the production and hadronization of quarkonium states in the hot QCD medium.  Preliminary measurements at forward rapidity and $2 < p_T < 6$~GeV by ALICE show that the polarization is independent of centrality \cite{ALICE_pol_centrality}. The $J/\psi$ polarization thus appears to be system independent even though the yields are quite sensitive to the presence or absence of hot or cold nuclear matter.

This assertion could be further tested by extending the $p_T$ range of the Pb+Pb polarization data to $p_T > 10$~GeV, where regeneration is no longer important and only $J/\psi$ suppression plays a role.  This higher $p_T$ region is also outside the range where other low $p_T$ effects such as $p_T$ broadening are also less important.  While much more difficult due to the strong suppression, a measure of $\psi$(2S) polarization in Pb+Pb collisions at high $p_T$ could also provide an independent check.


\section*{Acknowledgements}
R.V. thanks N. Brambilla, A. Vairo, H.-S. Chung, and E. Scomparin for discussions.  She also thanks the Technical University of Munich, where this work was initially originated, for hospitality as an August-Wilhelm Scheer Visiting Professor.  This work was supported in part by the U.S. Department of Energy by Lawrence Livermore National Laboratory under Contract No. DE-AC52-07NA27344, the U.S. Department of Energy, Office of Science, Office of Nuclear Physics (Nuclear Theory) under Contract No. DE-SC-0004014, and the LLNL LDRD program Grant No. LW-21-034.


\end{document}